\documentclass[twocolumn,showpacs,aps,prb,preprintnumbers,amsmath,amssymb]{revtex4}

\usepackage{graphicx}
\usepackage{dcolumn}
\usepackage{bm}
\usepackage{epsfig}

\begin{document}
\date{\today}

\title{
Transition between ordinary and topological insulator regimes \\ in two-dimensional resonant magnetotransport
}
\author{ G. Tkachov and E. M. Hankiewicz }
\affiliation{Institut f\"ur Theoretische Physik und Astrophysik,
Universit\"at W\"urzburg, Am Hubland, 97074 W\"urzburg, Germany}

\begin{abstract}
In the two-dimensional case the transition between ordinary and topological insulator states
can be described by a massive Dirac model with the mass term changing its sign at the transition point.
We theoretically investigate how such a transition manifests itself in resonant transport via
localized helical edge states. The resonance occurs in the middle of the band gap
due to a zero edge-state mode which is protected by the time-reversal symmetry,
also when coupled to the conducting leads.
We obtain the explicit dependence of the resonant conductance on the mass parameter
and an external magnetic field. The proposal may be of practical use, allowing one to determine
the orbital g-factor of helical edge states in two-dimensional topological insulators.
\end{abstract}

\maketitle
\section{Introduction}
Two- (2D) and three-dimensional (3D) topological insulators (TIs) are in the focus of current research
(see, recent reviews~\cite{Koenig08,Hasan10,Qi10} and references therein).
They differ from ordinary band insulators by the presence of protected edge (in 2D) or surface (in 3D) states
that support electric current and, for that reason, may have some application potential.
The 2D TIs are of special importance because
they exhibit the quantum spin Hall (QSH) state~\cite{Kane05,Bernevig06,Koenig07,Roth09}
that, unlike the quantum Hall (QH) states, does not require any magnetic field
and is characterized by time-reversal invariant gapless edge modes, while the bulk states are fully gapped.
Such edge modes, called sometimes "helical", consist of a pair of channels that propagate in the opposite directions
for opposite spin directions on the same edge.
Another interesting feature of the 2D TIs, which distinguishes them from 3D TIs, is that their low-energy physics
can be described by a massive Dirac model.~\cite{Bernevig06,Koenig07} In particular, the transition between the ordinary and topological insulating regimes corresponds to inversion of the effective relativistic mass of the Dirac-like fermions. At the transition the system behaves as a gas of massless Dirac fermions.~\cite{Buettner10}

The first 2D TI was realized experimentally in HgTe quantum wells.~\cite{Koenig07,Koenig08,Roth09}
In these experiments, the QSH regime
was detected by measuring the longitudinal conductance of two spin channels
propagating in the same direction on opposite edges of the sample.
This finding was further substantiated by the observed suppression of
the edge transport in a magnetic field,~\cite{Koenig07,Koenig08}
which breaks the time-reversal symmetry of the QSH state,
thus revealing the helical nature of edge channels.
Theoretically, the transport in QSH insulators has been analyzed
in both disordered~\cite{Maciejko09,Groth09,Ostrovsky10,GT11} and ballistic~\cite{Niu08,Yoko0910,Schmidt09,Novik10,GT10} regimes.

\begin{figure}[t]
\includegraphics[width=80mm]{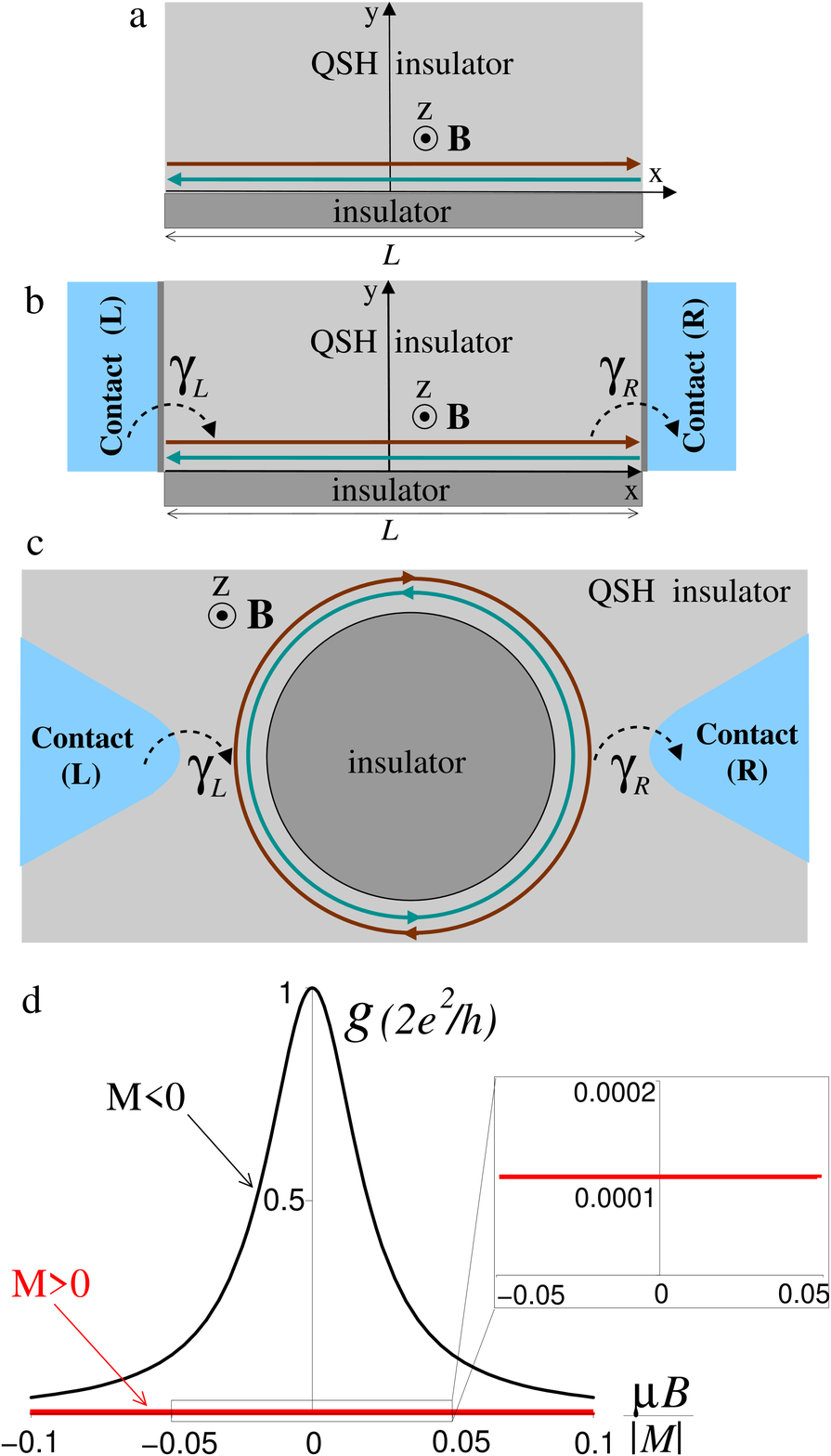}
\caption{
(a) Counter-propagating (helical) edge states at one of the boundaries ($y=0$) of a QSH insulator.
(b) and (c) Schematics of resonant tunneling via localized edge states on finite-length linear and circular boundaries in a QSH insulator. 
In (c) the edge states appear on the outer boundary of an island (e.g. suitably chosen ordinary insulator or a hole) surrounded by a QSH insulator.
$\gamma_L/\hbar$ and $\gamma_R/\hbar$ are tunneling rates between the edge and the left (L) and right (R) contacts.
(d) Magnetoconductance (\ref{g1_ZM}) for topologically nontrivial ($M<0$) and ordinary ($M>0$)
insulator states for $\gamma_L=\gamma_R=0.02\, |M|$. Inset: Enlarged view of $g(B)$ for $M>0$.
}
\label{Geo}
\end{figure}

In the present paper, we discuss another transport regime - resonant tunneling - as an alternative means of
probing helical edge states in 2D TIs. We consider edge states localized on a finite-length boundary
coupled by tunneling to two metallic leads (see also Fig.~\ref{Geo}).
The operation of such a device relies on the fact that gapless helical edge states are protected by the time-reversal
symmetry.~\cite{Kane05,Bernevig06} Therefore, for any tunneling coupling that preserves the time-reversal invariance
the edge spectrum contains a doubly degenerate zero mode in the middle of the
band gap, which provides the resonant level for tunneling between the leads.
If, upon the gap inversion, the systems goes into the ordinary insulating regime, the zero mode disappears
and the transport through the device is suppressed. This can be described by the following formula for the tunneling
conductance:
\begin{eqnarray}
g(M,B)&=&\frac{e^2}{h}
\left[
\frac{\gamma_R\gamma_L}{  ( M + |M|  + \mu B )^2  + \left(\frac{\gamma_R + \gamma_L}{2}\right)^2  }
+
\right.
\nonumber\\
&+&
\left.
\frac{\gamma_R\gamma_L}{  (M + |M|  - \mu B )^2  +  \left(\frac{\gamma_R + \gamma_L}{2}\right)^2 }
\right],
\label{g1_ZM}
\end{eqnarray}
where $M$ is the mass (gap) in the Dirac model~\cite{Bernevig06} and
$\gamma_L/\hbar$ and $\gamma_R/\hbar$ are the tunneling rates between the edge and the left (L) and right (R) contacts.
The two terms in Eq.~(\ref{g1_ZM}) arise from lifting the Kramers degeneracy of the zero mode in
an external magnetic field $B$, which involves the effective orbital magnetic moment of the edge states
$\mu = |e|\hbar \upsilon^2/2c|M|$ ($\upsilon$ is their velocity).

The transition between the topological and ordinary insulator states
manifests itself in the strong dependence of conductance (\ref{g1_ZM}) on
the sign of $M$ (see, also, Fig.~\ref{Geo}d). In the topologically nontrivial state with $M<0$,
the $B$-field dependence of conductance~(\ref{g1_ZM}) has a resonant (Lorenzian) line-shape around $B=0$,
while in the ordinary insulator regime with $M>0$ the conductance is small and field-independent,
$g \approx (e^2/h) \gamma_R\gamma_L/2M^2$ for $\gamma_{L,R} \ll |M|$ and $\mu B\ll 2|M|$.

Apart from the principal possibility to detect the edge zero mode, the resonant magnetoconductance~(\ref{g1_ZM})
can be used for determining the effective orbital g-factor of the helical edge states in a given material.
Such a need exists, e.g., in magnetotransport studies of HgTe quantum wells.~\cite{Koenig08}
For the typical velocity in HgTe wells, $\upsilon=5.5 \times 10^{5} $ms$^{-1}$,~\cite{Koenig08}
we estimate $\mu$ (in units of Bohr magneton $\mu_B$) as $\mu/\mu_B \approx 1700\, {\rm meV}/|M|$,
where $M$ is measured in meV. For the realizable values of $M\sim 10$ meV, the orbital g-factor is quite large
$\sim 170$.

The subsequent sections describe the details of our theoretical analysis.
In Sec.~\ref{QSHI} we solve the boundary problem for helical edge states in a QSH insulator subject
to a finite magnetic field. Section~\ref{transport} presents a more general formula for the resonant magnetoconductance in finite magnetic fields, from which we derive Eq.~(\ref{g1_ZM}), and finally Sec.~\ref{conclusions} concludes the paper.

\section{Quantum spin Hall insulator in a magnetic field}
\label{QSHI}

\subsection{Boundary problem}
We begin by analyzing the edge states on a boundary of a QSH insulator of finite length $L$,
imposing symmetric boundary conditions at the ends $x=\pm L/2$ (see, Fig.~\ref{Geo}a).
The symmetric boundary conditions result in a vanishing particle current density at $x=\pm L/2$,
thus providing a confinement for Dirac fermions in a QSH insulator (see also Appendix \ref{conf}).
For concreteness, we use the effective 4-band model derived in Ref.~\onlinecite{Bernevig06} for HgTe quantum wells.
In this approach one works in the basis of the four states near the $\Gamma$  (${\bf p}=0$)
point of the Brillouin zone:
$|e_1 +\rangle$, $|h_1 +\rangle$, $|e_1 -\rangle$, and $|h_1 -\rangle$,
where $e_1$ and $h_1$ are the s-like electron and p-like hole QW subbands, respectively.
The index $\tau=\pm$ accounts for the spin degree of freedom.
The effective two-dimensional Hamiltonian can be approximated by a diagonal matrix in ${\tau}$ space~\cite{Bernevig06}:
\begin{eqnarray}
H=
\left(
  \begin{matrix}
    h_{\bf p} & 0 \\
    0 & h^{\ast}_{\bf -p} \\
  \end{matrix}
\right),
h_{\bf p} = {\bf d}_{\bf p}\mbox{\boldmath$\sigma$},
\,
{\bf d}_{\bf p} =(\upsilon p_x, -\upsilon p_y, M).
\label{Heff}
\end{eqnarray}
where Pauli matrices $\sigma_{x,y,z}$ act in subband space,
$\upsilon\approx 5.5 \times 10^{5} $ms$^{-1}$ is the effective velocity~\cite{Koenig08},
and $M$ determines the bulk band gap at ${\bf p}=0$.
In Eq.~(\ref{Heff}) we omit terms $\propto {\bf p}^2$
since our main results will hold for the vicinity of the ${\bf p}=0$ point.
Up to a unitary transformation, Eq.~(\ref{Heff}) is equivalent to a massive Dirac Hamiltonian
\begin{equation}
H_D = \upsilon \tau_z\mbox{\boldmath$\sigma$}{\bf p} + M\tau_z\sigma_z,
\label{H_D}
\end{equation}
$\tau_z$ is the Pauli matrix in spin space.
We will work with the corresponding retarded Green's function defined by
\begin{eqnarray}
[\epsilon\, I  - H_D]{\hat G}({\bf r},{\bf r}^\prime)=I \delta( {\bf r}-{\bf r}^\prime ),
\label{Eq_G}
\end{eqnarray}
where
\begin{equation}
{\bf p}=-\hbar i{\bf\nabla}  - e{\bf A}({\bf r})/c,\quad
{\bf A}({\bf r})=(-By,0,0).
\label{p}
\end{equation}
${\bf A}({\bf r})$ is the vector potential of an external magnetic field $B$,
and $I=\tau_0\sigma_0={\rm diag}(1,1,1,1)$.

The system size in the $y$ direction is considered much larger than the characteristic
decay length of the edge states, $\sim \hbar \upsilon/|M|$,
allowing us to focus on one of the edges ($y=0$ in Fig.~\ref{Geo}a).
At $y=0$ we impose the condition,~\cite{Berry87}
\begin{eqnarray}
G({\bf r},{\bf r}^\prime)|_{y=0}=
\tau_0\otimes\sigma_x\, G({\bf r},{\bf r}^\prime)|_{y=0}.
\label{BC}
\end{eqnarray}
It can be obtained by introducing a large mass term ($M \to +\infty$)
outside the physical area of the system. Therefore, with $M<0$ in the bulk of the system
the boundary condition (\ref{BC}) is equivalent to a mass domain wall which
is responsible for the appearance of the edge states in a 2D QSH insulator in close analogy with
the 3D situation considered in Ref.~\onlinecite{Volkov85}.
One can check that the boundary condition (\ref{BC}) indeed ensures vanishing of the normal component
of the particle current:
\begin{eqnarray}
j_y(x,0)&=&\psi^\dagger(x,0) \tau_3\otimes\sigma_y \psi(x,0)
\label{current}\\
&=&
\psi^\dagger(x,0)\tau_0\otimes\sigma_x
(\tau_3\otimes\sigma_y)
\tau_0\otimes\sigma_x\psi(x,0)
\nonumber\\
&=&-\psi^\dagger(x,0) \tau_3\otimes\sigma_y \psi(x,0)=-j_y(x,0)=0,
\nonumber
\end{eqnarray}
where we switched to the creation and annihilation operators subject to boundary condition (\ref{BC}):
$\psi(x,0)=\tau_0\otimes\sigma_x\,\psi(x,0)$ and $\psi^\dagger(x,0)=\psi^\dagger(x,0)\tau_0\otimes\sigma_x$.

\subsection{Green's function solution}
The solution of this boundary problem can be obtained, following the same steps as in the problem of the edge states in graphene.~\cite{GT0907,Madrid} The Green's function is block-diagonal in Kramers partner ($\tau$) space:
\begin{eqnarray}
\hat{G}=
\left(
\begin{array}{cc}
\hat{G}_+ & 0\\
  0 & \hat{G}_-
\end{array}
\right),\,
\hat{G}_\tau=
\left(
\begin{array}{cc}
G_{11|\tau} & G_{12|\tau}\\
G_{21|\tau} & G_{22|\tau}
\end{array}
\right),\,
\tau = \pm,
\end{eqnarray}
where $\hat{G}_\tau$ is a matrix in space of the two QW subbands (or, generally, in space of the upper and lower components of the Weyl spinor), labeled by indices 1 and 2.
Writing Eq.~(\ref{Eq_G}) in components, it is easy to express the off-diagonal elements $G_{12|\tau}$ and $G_{21|\tau}$
in terms of the diagonal ones as follows
\begin{eqnarray}
\hat{G}_\tau=\left(
\begin{array}{cc}
G_{11|\tau} & \frac{\upsilon p_-}{\tau\epsilon - M}\, G_{22|\tau}\\
\frac{\upsilon p_+}{\tau\epsilon + M}\, G_{11|\tau} & G_{22|\tau}
\end{array}
\right),
\,
p_\pm = p_x \pm ip_y,
\label{G_diag}
\end{eqnarray}
where for $G_{11|\tau}$ and $G_{22|\tau}$ we have
\begin{eqnarray*}
&
[\epsilon^2 - M^2  - \upsilon^2 p_-p_+ ]G_{11|\tau}=
(\epsilon + M\tau)\delta( {\bf r}-{\bf r}^\prime ),
&
\\
&
[\epsilon^2 - M^2  - \upsilon^2 p_+ p_- ]G_{22|\tau}=
(\epsilon - M\tau)\delta( {\bf r}-{\bf r}^\prime ).
&
\end{eqnarray*}
Expanding
\begin{eqnarray*}
\hat{G}_{\tau}({\bf r},{\bf r}^\prime)=\sum_{n=-\infty}^{\infty} G_{\tau n}(y,y^\prime)\,e^{ik_n(x-x^\prime)}/L,
\quad
k_n=2\pi n/L,
\end{eqnarray*}
we obtain ordinary differential equations for $G_{11|\tau n}(y,y^\prime)$ and $G_{22|\tau n}(y,y^\prime)$:
\begin{eqnarray}
&&
\left[\frac{\partial^2}{\partial Y^2} - \frac{(Y-Y_0)^2}{4} - a_1 \right]G_{11|\tau n}=
\frac{ \epsilon + M\tau}{\hbar^2\upsilon^2/\lambda}\delta( Y-Y^\prime ),
\nonumber\\
&&
a_1=\frac{M^2 - \epsilon^2}{(\hbar \upsilon/\lambda)^2} - \frac{{\rm sgn} (eB)}{2},
\label{Eq1}
\end{eqnarray}
\begin{eqnarray}
&&
\left[\frac{\partial^2}{\partial Y^2} - \frac{(Y-Y_0)^2}{4} - a_2 \right]G_{22|\tau n}=
\frac{ \epsilon - M\tau }{\hbar^2\upsilon^2/\lambda}\delta( Y-Y^\prime ),
\nonumber\\
&&
a_2=\frac{M^2 - \epsilon^2}{ (\hbar \upsilon/\lambda)^2 } + \frac{{\rm sgn} (eB)}{2},
\label{Eq2}
\end{eqnarray}
where position $Y=y/\lambda$ is measured in units of magnetic length $\lambda=(c\hbar/2|eB|)^{1/2}$,
and $Y_0=-2\lambda\, k_n\, {\rm sgn} (eB)$ is the center of the oscillator.
The boundary conditions for Eqs.~(\ref{Eq1}) and (\ref{Eq2}) are derived from Eq.~(\ref{BC})
by writing it in components and using Eq.~(\ref{G_diag}):
\begin{eqnarray}
&&
\left.\frac{\partial G_{11|\tau n}(Y,Y^\prime) }{\partial Y}\right|_{Y=0}=\kappa_{1}\, G_{11|\tau n}(0,Y^\prime),
\label{BC1}\\
&&
\kappa_1=\frac{\tau \varepsilon + M}{\hbar \upsilon/\lambda} + \frac{Y_0 \, {\rm sgn} (eB)}{2},
\label{k1}\\
%
&&
\left.\frac{\partial G_{22|\tau n}(Y,Y^\prime) }{\partial Y}\right|_{Y=0}=\kappa_{2}\, G_{22|\tau n}(0,Y^\prime),
\label{BC2}\\
&&
\kappa_2=-\frac{\tau \varepsilon - M}{\hbar \upsilon/\lambda} - \frac{Y_0 \, {\rm sgn} (eB)}{2}.
\label{k2}
\end{eqnarray}

We seek the solution to Eq.~(\ref{Eq1}) in the form of the linear combination:
\begin{eqnarray}
G_{11|\tau n}(Y,Y^\prime)=G^\infty_{11|\tau n}(Y,Y^\prime) + A_1 U(a_1,Y-Y_0).
\label{Sol}
\end{eqnarray}
The first term is the Green's function of the unbounded system (source term),
\begin{eqnarray}
&&
G^\infty_{11|\tau n}=C_1
\left\{
\begin{array}{c}
U(a_1,Y-Y_0)U(a_1,-Y^\prime+Y_0)|_{Y\geq Y^\prime}, \\
U(a_1,Y^\prime-Y_0)U(a_1,-Y+Y_0)|_{Y^\prime\geq Y},
\end{array}
\right.
\nonumber\\
&&
C_1=-\frac{ \lambda( \epsilon + M\tau )\Gamma(a_1 +1/2) }{ \sqrt{2\pi}\hbar^2\upsilon^2 },
\label{G1_infty}
\end{eqnarray}
where $U(a_1,Y-Y_0)$ is the parabolic cylinder function (see, e.g. Ref.~\onlinecite{AS})
and $\Gamma( a_1 +1/2 )$ is Euler's gamma function.
The second term in Eq.~(\ref{Sol}) is the solution of the corresponding homogeneous equation, decaying for $Y\to\infty$. The coefficient $A_1$ can be found from boundary condition (\ref{BC1}),
which finally yields
\begin{widetext}
\begin{eqnarray}
G_{11|\tau n}(Y,Y^\prime)=G^{\infty}_{11|\tau n}(Y,Y^\prime) -
C_1\frac{ (Y_0/2 + \kappa_1) U( a_1 , Y_0 ) - U( a_1-1 , Y_0 )}
        { (Y_0/2 + \kappa_1) U( a_1 , Y_0 ) + U( a_1-1 , Y_0 ) }
                        \,U( a_1 , Y - Y_0 )U( a_1 , Y^\prime - Y_0 ).
\label{G_11}
\end{eqnarray}
We have also used the recurrence relation $U^\prime(a,x)=(x/2)U(a,x) - U(a-1,x)$.~\cite{AS}
$G_{22|\tau n}(Y,Y^\prime)$ is obtained similarly and has the form:
\begin{eqnarray}
G_{22|\tau n}(Y,Y^\prime)=G^{\infty}_{22|\tau n}(Y,Y^\prime) -
C_2\frac{ (Y_0/2 + \kappa_2) U( a_2 , Y_0 ) - U( a_2-1 , Y_0 )}
        { (Y_0/2 + \kappa_2) U( a_2 , Y_0 ) + U( a_2-1 , Y_0 )}
                        \,U( a_2 , Y - Y_0 )U( a_2 , Y^\prime - Y_0 ),
\label{G_22}
\end{eqnarray}
\end{widetext}
with the source term
\begin{eqnarray}
&&
G^\infty_{22|\tau n}=C_2
\left\{
\begin{array}{c}
U(a_2,Y-Y_0)U(a_2,-Y^\prime+Y_0)|_{Y\geq Y^\prime}, \\
U(a_2,Y^\prime-Y_0)U(a_2,-Y+Y_0)|_{Y^\prime\geq Y},
\end{array}
\right.
\nonumber\\
&&
C_2=-\frac{ \lambda( \epsilon - M\tau )\Gamma(a_2 +1/2) }{ \sqrt{2\pi}\hbar^2\upsilon^2 }.
\label{G2_infty}
\end{eqnarray}

\subsection{ Transition between ordinary and topological regimes in weak magnetic fields }

Let us now demonstrate how the Green's function formalism describes
the transition between ordinary and topological regimes.
We will assume that energy $\epsilon$ and magnetic field $B$ are small in the sense that
\begin{eqnarray}
|\epsilon|\ll |M|,\quad \hbar\upsilon/\lambda\ll |M|.
\end{eqnarray}
This corresponds to the positive and large $a_1$ [see, Eq.~(\ref{Eq1})],
when the parabolic cylinder functions in Eq.~(\ref{G_11}) assume the form
$U(a,x)\approx \sqrt{\pi}/( 2^{a/2 + 1/4} \Gamma(3/4 + a/2) ){\rm e}^{-\sqrt{a} \, x}$.~\cite{AS}
Using then Stirling's expansion for $\Gamma(x)$, we find
\begin{eqnarray}
&&
G_{11|\tau n}(y,y^\prime)\approx G^{\infty}_{11|\tau n}(y,y^\prime) +
\label{G_11weak}\\
&&
+
\frac{ \epsilon + \tau M }{2 \hbar \upsilon |M|}
\frac{  M - |M| - \frac{h\upsilon}{L}\, n + \mu B + \tau\epsilon }
     {  M + |M| - \frac{h\upsilon}{L}\, n - \mu B + \tau\epsilon }
{\rm e}^{ -|M|(y+y^\prime)/\hbar \upsilon }.
\nonumber
\end{eqnarray}
The nontrivial dependence of the edge (second) term on the mass sign describes the transition between the ordinary and topological insulator states. Indeed, the gapless edge modes appear only when $M$ reverses its sign from positive to negative:
\begin{eqnarray}
\epsilon_{\tau , n} =\frac{hv}{L}\,\tau n + \mu \tau B,\quad M<0.
\label{E}
\end{eqnarray}
Unlike the conventional quantum Hall systems~\cite{Halperin82} such edge states exist in a zero or weak magnetic field, when their localization length $\hbar\upsilon/|M|$ is determined by the band gap $|M|$ rather than the magnetic field $B$.~\cite{strong}  The weak magnetic field results in a Zeeman-like term $\mu \tau B$ in Eq. (\ref{E}) where
the effective magnetic moment $\mu=|e|\hbar \upsilon^2/2c|M|$ has an orbital origin: it is related to the magnetic flux
through the finite width $\hbar\upsilon/|M|$ of the edge state. The magnetic field lifts the Kramers degeneracy $\epsilon_{-\tau,-n}=\epsilon_{\tau,n}$, splitting the Kramers partners by $2\mu B$ in energy.

\begin{figure}[b]
\includegraphics[width=80mm]{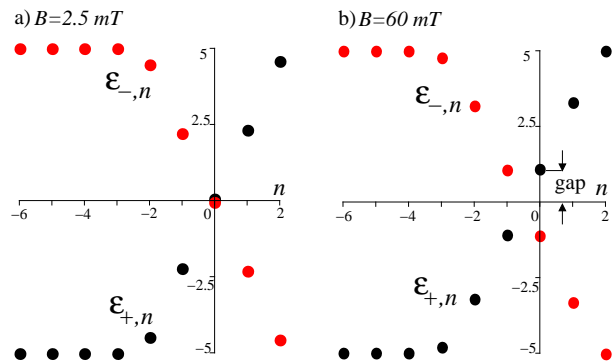}
\caption{
Energies of spin-up $\epsilon_{+,n}$ and spin-down $\epsilon_{-,n}$ edge states (in meV)
vs. wave number $n$ for $M=-5$ meV, $\upsilon=5.5 \times 10^{5} $ms$^{-1}$ and $L=1\, \mu$m.
(a) In a vanishing magnetic field the spectrum is gapless with the Kramers-degenerate zero mode at $n=0$.
(b) The degeneracy is lifted in a finite field opening a gap at $n=0$.
For $B=60$ mT the gap energy is approximately $1$ meV.
}
\label{Spectrum}
\end{figure}

In the general case, the analytical results for the edge-state spectrum  are corroborated by the numerical solution
of equation $(Y_0/2 + \kappa_1) U( a_1 , Y_0 ) + U( a_1-1 , Y_0 )=0$, which corresponds to the pole
of Green's function (\ref{G_11}). For positive $M$ we again find no subgap states,
while for $M<0$ the solution is shown in Fig.~\ref{Spectrum}.
In a vanishing magnetic field (see, Fig.~\ref{Spectrum}a),
the spectrum contains a Kramers-degenerate zero mode with $n=0$ in the middle of the band gap.
The finite field lifts the Kramers degeneracy, opening a gap at $n=0$ (see, Fig.~\ref{Spectrum}b).
We emphasize that the gap opening is the consequence of the discreteness of the spectrum.
For the continuum spectrum (i.e. for $L\to\infty$) the effect of the orbital magnetic field
reduces to a shift of the crossing point of the $\tau=\pm$ branches.~\cite{GT10}

\subsection{0D case}

Hereafter we focus on the 0D case which admits a simpler analytical treatment at finite
magnetic fields. This limit is realized when the level spacing $h\upsilon/L$ is much larger than both
thermal activation energy $k_B T$ and level broadening $\gamma$ (due to the coupling to metallic leads, see Sec.~\ref{transport}):
\begin{equation}
h \upsilon/L > k_B T,\, \gamma.
\label{0D}
\end{equation}
Under such conditions, the transport at energy $\epsilon=0$ is determined by the presence (or absence)
of the Kramers-degenerate zero mode with $n=0$ (see, Fig.~\ref{Spectrum}).
Consequently, in Green's functions~(\ref{G_11}) and (\ref{G_22}) one can set $Y_0=0$ and use
$U(a,0)=\sqrt{\pi}/2^{a/2 +1/4}\Gamma( 3/4 + a/2)$ (see, Ref.~\onlinecite{AS}), which yields
\begin{widetext}
\begin{eqnarray}
G_{11|\tau 0}(Y,Y^\prime)\approx
-C_1\frac{ \lambda (\epsilon + M\tau) B( 1/2 , 1/4 + a_1/2 ) - \sqrt{2\pi}\hbar\upsilon\tau }
         { \lambda (\epsilon + M\tau) B( 1/2 , 1/4 + a_1/2 ) + \sqrt{2\pi}\hbar\upsilon\tau  }
                        \,U( a_1 , Y  )U( a_1 , Y^\prime ),
\label{G0_11}
\end{eqnarray}
\begin{eqnarray}
G_{22|\tau 0}(Y,Y^\prime) \approx
-C_2\frac{ \lambda (\epsilon - M\tau) B( 1/2 , 1/4 + a_2/2 ) + \sqrt{2\pi}\hbar\upsilon\tau }
         { \lambda (\epsilon - M\tau) B( 1/2 , 1/4 + a_2/2 ) - \sqrt{2\pi}\hbar\upsilon\tau }
                        \,U( a_2 , Y  )U( a_2 , Y^\prime ),
\label{G0_22}
\end{eqnarray}
\end{widetext}
where $B( 1/2 , 1/4 + a/2 )$ is Euler's beta function (we have omitted the bulk terms since they do not have the poles
inside the band gap). Next, we integrate Eqs.~(\ref{G0_11}) and (\ref{G0_22}) over the remaining variable $y$
and introduce the 0D functions as
$G_{11|\tau 0}(\epsilon)=\int^\infty_0 G_{11|\tau 0}(Y,Y)dY$ and
$G_{22|\tau 0}(\epsilon)=\int^\infty_0 G_{22|\tau 0}(Y,Y)dY$.
The integrals can be expressed in terms of the digamma, $\psi$, and gamma functions using the identity
$\int^\infty_0 U^2(Y)dY = \sqrt{\pi/2} [ \psi(3/4 + a/2)  - \psi(1/4 +a/2) ]/2\Gamma(1/2 +a) $ (see, Ref.~\onlinecite{BE}):
\begin{widetext}
\begin{eqnarray}
G_{11|\tau 0}(\epsilon)\approx
\frac{ \lambda^2 (\epsilon + M\tau)[ \psi(3/4 + a_1/2)  - \psi(1/4 +a_1/2) ] }
     { 2\pi\hbar^2\upsilon^2}
\frac{ \lambda (\epsilon + M\tau) B( 1/2 , 1/4 + a_1/2 ) - \sqrt{2\pi}\hbar\upsilon\tau }
     { \lambda (\epsilon + M\tau) B( 1/2 , 1/4 + a_1/2 ) + \sqrt{2\pi}\hbar\upsilon\tau },
\label{G00_11}
\end{eqnarray}
%
\begin{eqnarray}
G_{22|\tau 0}(\epsilon)\approx
\frac{ \lambda ^2(\epsilon - M\tau)[ \psi(3/4 + a_2/2)  - \psi(1/4 +a_2/2) ] }
     { 2\pi\hbar^2\upsilon^2}
\frac{ \lambda (\epsilon - M\tau) B( 1/2 , 1/4 + a_2/2 ) + \sqrt{2\pi}\hbar\upsilon\tau }
     { \lambda (\epsilon - M\tau) B( 1/2 , 1/4 + a_2/2 ) - \sqrt{2\pi}\hbar\upsilon\tau }.
\label{G00_22}
\end{eqnarray}
\end{widetext}
In the next section we will use these equations to calculate the resonant magnetoconductance through
the zero edge mode.

\begin{figure}[b]
\includegraphics[width=70mm]{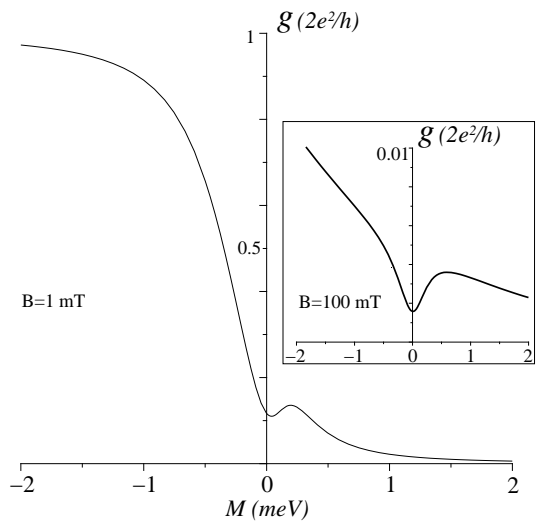}
\caption{Conductance (\ref{g_ZM}) in units of $2e^2/h$ versus gap $M$ (meV) for $B=1$ mT.
Inset shows the same dependence for higher field $B=100$ mT; $\upsilon=5.5 \times 10^{5} $ms$^{-1}$ and $\gamma_L=\gamma_R=0.3$ meV. }
\label{G_M}
\end{figure}

\section{Resonant magnetotransport}
\label{transport}

Transport through the localized helical edge states can be realized in the geometries shown in Figs. \ref{Geo}b and c \cite{Circle} where the edge is coupled to two metallic leads (L and R) via tunneling.
Following the general approach of Ref.~\onlinecite{Meir92} we describe the coupling by the tunneling Hamiltonian,

\begin{equation}
H_T= \sum_{ {\bf k},\alpha\in L,R}\sum_{n,\tau}
( V_{{\bf k}\alpha, n\tau} c^\dagger_{{\bf k}\alpha} d_{n\tau} + H.C.),
\label{H_T}
\end{equation}
where operator $c^\dagger_{{\bf k}\alpha}$ ($c_{{\bf k}\alpha}$) creates (destroys) an electron with momentum
${\bf k}$ and spin $\alpha$ in lead L or R, and $d^\dagger_{n\tau}$ and $d_{n\tau}$ are the creation and annihilation operators for a localized edge state with quantum numbers $n$ and $\tau$.
The tunneling matrix elements $V_{{\bf k}\alpha, n\tau}$ are assumed spin-independent.
Further, we will work in the 0D limit (\ref{0D}) where only the zero edge mode, described by $d^\dagger_{0\tau}$ and $d_{0\tau}$ in Eq.~(\ref{H_T}), needs to be taken into account.
Since the problem is similar to the tunneling through a resonant level,
we can apply the known formula for the zero-temperature two-terminal conductance:~\cite{Meir92}
\begin{eqnarray}
g=-\frac{e^2}{h}\frac{2\gamma_R\gamma_L}{\gamma_R+\gamma_L}{\rm Tr \, Im}\,{\cal\hat G}(\epsilon=0),
\label{g}
\end{eqnarray}
where the tunneling energies $\gamma_{L,R}$ are expressed in terms of the tunneling matrix elements
and Green's functions of the leads.~\cite{Meir92}
In Eq.~(\ref{g}) ${\rm Tr \, Im}$ denotes the trace in $\tau\otimes\sigma$ space of the imaginary part of the retarded Green's function ${\cal\hat G}(\epsilon)$, which is related to the Green's function ${\hat G}(\epsilon)$ calculated above through the Dyson equation,
${\cal\hat G}(\epsilon)=\hat{G}(\epsilon) + \hat{G}(\epsilon) \hat{\Sigma} {\cal\hat G}(\epsilon)$,
involving the tunneling self-energy $\hat{\Sigma}$.
For metallic leads and in the absence of spin scattering, $\hat{\Sigma}$ is energy-independent
and diagonal in $\tau\otimes\sigma$ space:~\cite{Meir92}
\begin{equation}
\Sigma=-i\gamma\, I, \quad \gamma=(\gamma_R+\gamma_L)/2.
\end{equation}
The solution of the Dyson equation takes the form
${\cal\hat G}(\epsilon)=\hat{G}(\epsilon + i\gamma)$.
This self-consistently accounts for the edge-spectrum broadening due to the presence of the leads.
With the 0D Green's functions (\ref{G00_11}) and (\ref{G00_22}), Eq.~(\ref{g}) reads
\begin{eqnarray}
g=-\frac{e^2}{h}\frac{2\gamma_R\gamma_L}{\gamma_R+\gamma_L}
\sum_{\tau=\pm 1} {\rm Im}
[ G_{11|\tau 0}(i\gamma)+G_{22|\tau 0}(i\gamma) ],
\label{g1}
\end{eqnarray}
or, explicitly,
\begin{eqnarray}
&&
g(M,B)=\frac{2e^2}{h}
\left[
\frac{\Gamma_1(x)\Gamma^\prime_1(x)\,\gamma_R\gamma_L}{  ( M + |M| \Gamma_1(x)/\sqrt{x} )^2  + \gamma^2  }
+
\right.
\nonumber\\
&&+
\left.
\frac{\Gamma_2(x)\Gamma^\prime_2(x)\,\gamma_R\gamma_L}{  ( M + |M| \Gamma_2(x)/\sqrt{x} )^2  + \gamma^2  }
\right]_{x=B_{_M}/|B|},
\label{g_ZM}
\end{eqnarray}
where $B_{_M}=cM^2/(4 |e| \hbar \upsilon^2 )$ and the functions $\Gamma_{1,2}(x)$ and their derivatives $\Gamma^\prime_{1,2}(x)$ are expressed in terms of the gamma functions:
\begin{eqnarray}
\Gamma_1(x)=\Gamma(x+1/2)/\Gamma(x), \quad \Gamma_2(x)=x/\Gamma_1(x).
\label{Gamma_12}
\end{eqnarray}

Let us analyze the results for the magnetoconductance.
Equation~(\ref{g_ZM}) is valid for arbitrary ratio of $B$ and the characteristic field, $B_{_M}$,
related to the band gap $M$. Equation~(\ref{g1_ZM}), presented in the introduction, follows from Eq.~(\ref{g_ZM})
in the case of weak fields (or sufficiently large gap $|M|$) where $x^{-1}=|B|/B_{_M}\ll 1$.
To see this we use Stirling's formula to obtain the expansions
$\Gamma_1(x)\approx \sqrt{x} ( 1-x^{-1}/8 )$ and $\Gamma_2(x)\approx \sqrt{x} ( 1 + x^{-1}/8 )$
and then insert them into Eq.~(\ref{g_ZM}).

As $|M|$ becomes smaller (i.e. $x=B_{_M}/|B| \leq 1$ )
one should use Eq.~(\ref{g_ZM}) instead of (\ref{g1_ZM}).
Figure~\ref{G_M} shows the dependence $g(M)$ given by Eq.~(\ref{g_ZM}).
The conductance varies smoothly from $2e^2/h$ in the inverted regime ($M<0$)
to zero in the ordinary insulator state with $M>0$.
Interestingly, there is a local minimum at $M=0$.
To demonstrate this analytically, let us examine Eq.~(\ref{g_ZM}) for  $M\ll \gamma$
and  $x=B_{_M}/|B|\ll 1$.  Using asymptotics $\Gamma_1(x)\approx \pi^{1/2}x$
and $\Gamma_2(x)\approx \pi^{-1/2}$, we expand: $g(M)\approx g(0) + \Delta g(M)$,
where the correction $\Delta g(M)$ is given by
\begin{eqnarray}
\frac{\Delta g(M)}{e^2/\hbar}=\frac{\gamma_R\gamma_L}{\gamma^2}\,\frac{B_{_M}}{|B|}=
\frac{\gamma_R\gamma_L}{\gamma^2}\,\frac{cM^2}{4 \hbar \upsilon^2 |eB|}.
\label{Dg}
\end{eqnarray}
It is quadratic in $M$ reflecting the analyticity of the conductance in the limit $M\to 0$ at finite $B$ and $\gamma$.
Therefore, on both sides of the transition the conductance initially increases with $M$.
However, for $M>0$ the quadratic growth is followed by the suppression
when the ordinary insulator regime with the gap $M>\gamma$ is reached.

\section{Conclusions}
\label{conclusions}

We have proposed a model for resonant magnetotransport via localized helical edge states
in a quantum spin Hall system. To calculate the magnetoconductance we used the Green's function formalism
allowing us to treat the topological and ordinary insulator regimes on equal footing.
In the topologically nontrivial state, the magnetic-field dependence of the conductance has a resonant (Lorenzian) line-shape explicitly depending on the orbital magnetic moment of the helical edge states.
One could use this to extract the orbital g-factor of the helical edge states in a given material.
The necessary condition to test our predictions is a sufficiently large edge level spacing $h\upsilon/L$
(see, Eq.~(\ref{0D}) ). For HgTe wells with $\upsilon=5.5 \times 10^{5} $ms$^{-1}$ and $L=1$ $\mu$m the level spacing is  about $2.5$ meV (see, also, Fig.~\ref{Spectrum}), implying working temperatures up to a few kelvin.
Note that with increasing $B$ one should expect oscillations of the tunneling conductance that follow the zero-mode ($B=0$) resonance because the states with nonzero momentum [$n\not =0$ in Eq.~(\ref{E})] will periodically cross the zero energy. Similar oscillations are expected in the gate voltage dependence of the conductance, which can be used to adjust the position of the Fermi level in the band gap.

\begin{acknowledgements}
We thank S.-C. Zhang, Q.-L. Qi, B. Trauzettel, P. Recher and P. Michetti for helpful discussions.
This work was supported by DFG Grant HA5893/1-1.
\end{acknowledgements}


\appendix
\section{Electron confinement with symmetric boundary conditions}
\label{conf}

We consider a system of finite length $L$ (in the $x$ direction in Fig. \ref{Geo}a) subject to a perpendicular magnetic field ${\bf B}$ (\ref{p}) and described by the equations
\begin{eqnarray}
&& i\hbar \partial_t \psi_+({\bf r},t)=
\label{Eq_Psi+}\\
&&
=(
i\hbar\upsilon [\sigma_x (\partial_x + ieBy/c\hbar) + \sigma_y \partial_y ] - M\sigma_z
)
\psi_+({\bf r},t),
\nonumber\\
&& i\hbar \partial_t \psi_-({\bf r},t)=
\label{Eq_Psi-}\\
&&
=(
-i\hbar\upsilon [\sigma_x (\partial_x + ieBy/c\hbar) + \sigma_y \partial_y ] + M\sigma_z
)
\psi_-({\bf r},t),
\nonumber
\end{eqnarray}
for the operators $\psi_\pm({\bf r},t)$ of the Kramers-partner states of Hamiltonian (\ref{H_D}).
At the ends of the system we impose the symmetric boundary conditions:
\begin{equation}
\psi_\pm(-L/2,y)=\psi_\pm( L/2,y).
 \label{BC_symm}
\end{equation}

Our goal is to show that the symmetric boundary conditions (\ref{BC_symm}) yield vanishing current density,
\begin{equation}
j_x(\pm L/2,y)=0,
\label{j_x=0}
\end{equation}
and in this sense the particle is confined in the region $|x|\leq L/2$. This becomes possible due to the specific structure of the Dirac particle current density,
\begin{eqnarray}
j_x(x,y)=\psi^\dagger_+(x,y)\sigma_x \psi_+(x,y) - \psi^\dagger_-(x,y)\sigma_x \psi_-(x,y),\,\,\,\,
 \label{j_x}
\end{eqnarray}
and the parity symmetry of Eqs.~(\ref{Eq_Psi+}) and (\ref{Eq_Psi-}),
\begin{eqnarray}
\psi_-(x,y; B)=C \sigma_x \psi_+(-x,y; -B),\quad |C|^2=1.
 \label{parity}
\end{eqnarray}
We now use this relation to transform the second term in Eq. (\ref{j_x}) as follows
\begin{eqnarray}
j_x(x,y)&=& \psi^\dagger_+(x,y;B)\sigma_x \psi_+(x,y;B)
\label{j_x1}\\
&-& \psi^\dagger_+(-x,y;-B)\sigma_x \psi_+(-x,y;-B).
\nonumber
\end{eqnarray}
Setting $x=L/2$ (or $x=-L/2$) and using the boundary conditions (\ref{BC_symm}) we have
\begin{eqnarray}
j_x( \pm L/2,y)&=& \psi^\dagger_+(L/2,y;B)\sigma_x \psi_+(L/2,y;B)
\label{j_x2}\\
&-& \psi^\dagger_+(L/2,y;-B)\sigma_x \psi_+(L/2,y;-B).
\nonumber
\end{eqnarray}
For zero magnetic field, $B=0$, the two terms in Eq.~(\ref{j_x2}), i.e. the currents of the Kramers partners, cancel each other out, yielding vanishing current density (\ref{j_x=0}) at the ends of the system.
Moreover, the current density (\ref{j_x2}) should also vanish in a finite magnetic field, $B\not =0$, provided that the system is in the quantum spin Hall regime. Indeed, the current density (\ref{j_x2}) is {\em odd} in the magnetic field $B$ and, therefore, could only characterize the Hall response. However, in the quantum {\em spin} Hall regime considered in our paper a Kramers pair of counterpropagating edge states cannot generate the Hall response.


\begin{thebibliography}{00}
\bibitem{Circle}
In the case of the circular edge [Fig. \ref{Geo}(c)] we assume the circumference of the island, $L$, to be large enough to neglect the curvature effects, so that we can use the solution for the edge Green's function obtained in section \ref{QSHI}.
\end{thebibliography}
\end{document}